\documentclass[aps,pra,twocolumn,superscriptaddress,nofootinbib]{revtex4-2}

\usepackage{amsmath,amssymb,amsthm}
\usepackage{graphicx}
\usepackage{physics}
\usepackage{hyperref}
\usepackage{tikz-cd}
\usepackage{quantikz}
\newtheorem{theorem}{Theorem}

\def\BibTeX{{\rm B\kern-.05em{\sc i\kern-.025em b}\kern-.08em
    T\kern-.1667em\lower.7ex\hbox{E}\kern-.125emX}}
\newcommand{\TrOp}{\mathrm{Tr}}
\newcommand{\tn}[1]{\|#1\|}

\begin{document}

\title{P-divisibility conditions on subsystems in quantum dynamics}

\author{Anumita Mukhopadhyay}
\email{anumitamukherjee455@gmail.com}
\affiliation{Center for Quantum Engineering, Research and Education (CQuERE), TCG Centres for Research and Education in Science and Technology (TCG CREST), Kolkata 700091, India.}
\affiliation{Academy of Scientific and Innovative Research (AcSIR), Ghaziabad 201002, India.}
\author{Praggnyamita Ghosh}
\email{praggnyamitaghosh2003@gmail.com}
\affiliation{Ramakrishna Mission Vivekananda Educational and Research Institute, Belur Math, Howrah 711202, India.}
\affiliation{Center for Quantum Engineering, Research and Education (CQuERE), TCG Centres for Research and Education in Science and Technology (TCG CREST), Kolkata 700091, India.}
\author{Shibdas Roy}
\email{roy.shibdas@gmail.com}
\affiliation{Center for Quantum Engineering, Research and Education (CQuERE), TCG Centres for Research and Education in Science and Technology (TCG CREST), Kolkata 700091, India.}
\affiliation{Academy of Scientific and Innovative Research (AcSIR), Ghaziabad 201002, India.}

\begin{abstract}
We investigate the constraints imposed by global unitary
dynamics on the P-divisibility of local subsystems in a
bipartite system-environment setting.
Using the trace distance as a measure of state
distinguishability and exploiting its conservation under
unitary evolution, we develop a fully symmetric framework
that simultaneously tracks information flow in both the
open system $S$ and its environment $E$.
We first show that for initially uncorrelated states the
maximum net non-Markovian gain in one subsystem is
rigidly bounded by the initial distinguishability of the
other (Theorem~1).
We then establish a two-sided Correlation Window
(Theorem~2) that confines the sum of net changes in local
distinguishability: the upper edge is set by the total
initial marginal distinguishability, while the lower edge
is governed exactly by the dynamically generated bipartite
correlation norm, proving that any mutual loss of local
distinguishability is necessarily encoded into correlations.
Finally, we generalize both results to initially correlated
states (Theorem~3), showing that initial bipartite
correlations augment the capacity for non-Markovian gain
and that simultaneous P-indivisibility of both subsystems
is funded by the degradation of the initial correlation
reservoir.
These results are validated on four-qubit GHZ and W
circuit examples, demonstrating subsystem locking,
asymmetric single-subsystem backflow, and
correlation-funded simultaneous backflow.
\end{abstract}

\maketitle

\section{Introduction}

When a quantum system interacts with the surrounding environment then it is named as open quantum system. Noise is known to have adverse effect in quantum computing. The procedures to get rid of noise are also inefficient in terms of resources. Instead of bypassing the detrimental effects of noise, if it can be utilized in some manner, then Noisy Intermediate Scale Quantum (NISQ) era quantum computation can become significantly developed. In different researches, it is shown that certain kinds of noise can actually become helpful. Non-unitality and non-Markovianity are some of the conditions for the noise channels to be useful. Ref.~\cite{cirac,temporal,dissipaton,qcorr,noise} show how noise can enhance effectiveness of quantum computing. Non-Markovianity can also be a valuable resource in noise induced quantum computing \cite{srikanth,sibasish,Kumar,pathak,open_non_markov,teleport,naikoo}. The quantum channels acting on the system and the environment are generally completely positive trace preserving (CPTP) maps. When a map $\varepsilon(t_2,t_0)$ can be split into two intermediate CP maps, $\varepsilon(t_1,t_0)$ and $\varepsilon(t_2,t_1)$, representing the evolution for time $t_0 \leq t_1 \leq t_2$ then it is called Complete Positive (CP)-divisible map \cite{Rivas_2014,Utagi}. It is known that the interaction between system and environment introduces memory effect in the quantum system when they are CP-indivisible maps \cite{rivas,Rivas_2014}. In spite of that, sometimes CP-divisible maps also demonstrate memory effects \cite{banerjee,Utagi,Sbannerjee,srikanth,nonmarkov}. There are different measures to witness non-Markovianity as given in \cite{nonmarkov1,nonmarkov2, nonmarkov3,nonmarkov4,nonmarkov5,nonmarkov6,nonmarkov7,nonmarkov8,nonmarkov9}. The presence of memory effect can also be verified in terms of Positive (P)-indivisibility or BLP (Breuer-Laine-Piilo) criterion \cite{BLP}. In this measure, the trace distance between a pair of states is measured at the input and at the output of the quantum evolution. If there exists a pair of states that become more distinguishable after evolution, then it is said to be P-indivisible or a BLP non-Markovian channel. Such an increase of distinguishability points towards the presence of memory effect, even if the system and environment channels are CP-divisible. If the pair of states tends to be more indistinguishable over time then a flow of information from the system to the environment takes place. By contrast, an increase in distinguishability between a pair of states is interpreted as backflow of information from the environment to the system. To understand the behaviour of the noise channels, it is important to verify CP-divisibility \cite{ampgsr} as well as P-divisibility of the quantum systems. In this regard we have explored the condition of P-divisibility or BLP criterion in this work.
While the BLP measure effectively quantifies the backflow into a specific open system, the fundamental physical reality is a global unitary evolution acting on the joint system-environment Hilbert space. A critical development in this area was achieved by Laine, Piilo, and Breuer in 2010 \cite{Laine2010}, who demonstrated that the presence of initial system-environment correlations can fund an increase in the trace distance of the system above its initial value, violating the standard contraction property. Conversely, a temporary revival of distin-
guishability represents a backflow of information from the
environment to the principal system, signifying memory
effects and P-indivisible dynamics \cite{Chruscinski2011}.

Despite these advances, previous studies—including the seminal work of Laine \textit{et al.}\cite{Laine2010} have largely treated the system and environment asymmetrically. They typically focus solely on the distinguishability gain of the open system $S$, tracing out the environment $E$ immediately. The simultaneous dynamics of the environment, and how the global conservation of distinguishability dictates the trade-off between the two subsystems, has remained underexplored. 

In this paper, we present a fully symmetric, bipartite analysis of P-divisibility conditions. We bridge the gap between global unitary conservation and local non-Markovianity by deriving exact bounds on the net change in distinguishability for both subsystems simultaneously. In Sec.~\ref{sec:setup}, we define our setup and the fundamental unitary conservation laws. In Sec.~\ref{sec:initial correlation}, we analyze the asymmetric bounds established by previous literature and motivate the need for a bipartite approach. In Sec.~\ref{sec:prod}, we derive the strict capacity limits for net non-Markovian gain. In Sec.~\ref{sec:mid_corr}, we formally prove the Correlation Window, a rigorous bounding theorem that maps the trade-off between local non-Markovian backflow and the dynamical generation of bipartite correlations. In Sec.~\ref{sec:initial_corr}, we generalize our trade-off laws to accommodate initially correlated states, showing how initial correlations augment the capacity bounds. In Sec.~\ref{sec:example}, we illustrate these principles using concrete quantum circuit examples, demonstrating the phenomenon of subsystem locking. Finally, in Sec.~\ref{sec:discussion} we discuss our theorems in accordance with our numerical examples and we conclude by summarizing our work in Sec.~\ref{sec:conclusion}.

\section{Setup and Global Conservation}\label{sec:setup}

Consider a bipartite quantum system characterized by the composite Hilbert space $\mathcal{H}^{SE} = \mathcal{H}^S \otimes \mathcal{H}^E$, consisting of a system $S$ and an environment $E$. The composite system is assumed to be closed and undergoes unitary evolution governed by a global Hamiltonian $H$. Given two initial joint test states $\rho_1^{SE}(0)$ and $\rho_2^{SE}(0)$, their evolution is given by:
\begin{equation}
\rho_i^{SE}(t) = U(t)\rho_i^{SE}(0)U^\dagger(t), \quad i=1,2,
\end{equation}
where $U(t) = \exp(-iHt/\hbar)$. The reduced, locally accessible states at time $t$ are obtained via the partial trace:
\begin{align}
\rho_i^S(t) &= \Tr_E[\rho_i^{SE}(t)], \\
\rho_i^E(t) &= \Tr_S[\rho_i^{SE}(t)].
\end{align}

To quantify the distinguishability between the pairs of states, we utilize the trace distance, defined for any two density matrices as $D(\rho_1, \rho_2) = \frac{1}{2}||\rho_1 - \rho_2||_1$, where $||A||_1 = \Tr\sqrt{A^\dagger A}$ is the trace norm. The trace distance provides the maximum probability of  distinguishing between two states in a single measurement \cite{Nielsen2000}.

We define the trace distance between the two joint system-environment states and between the pair of subsystems for the initial states ($t=0$) and output states (at time $t$):
\begin{align}
\Delta_0^{SE} &= D(\rho_1^{SE}(0), \rho_2^{SE}(0)), \\
\Delta_t^{SE} &= D(\rho_1^{SE}(t), \rho_2^{SE}(t)),\\
D_S(0)&= D(\rho_1^S(0),\rho_2^S(0)),\\
D_S(t)&= D(\rho_1^S(t),\rho_2^S(t)),\\
D_E(0)&= D(\rho_1^E(0),\rho_2^E(0)),\\
D_E(t)&= D(\rho_1^E(t),\rho_2^E(t))
\end{align}

To construct our paper we have used some well-known tools as follows: 
\begin{itemize}
    \item The trace norm is multiplicative under the tensor products of the density operators~\cite{Bhatia1997}:
\begin{equation}
  \tn{A\otimes B}_1 = \tn{A}_1\cdot\tn{B}_1.
  \label{eq:mult}
\end{equation}
    \item The partial traces are CPTP maps and hence satisfy the data processing inequality~\cite{Lindblad1975,Ruskai1994}:
\begin{equation}
  \tn{\TrOp_E[A]}_1 \leq \tn{A}_1.
  \label{eq:dpi}
\end{equation}
\item The fundamental anchor of our analysis is the isometric nature of unitary conjugation. Because unitary operations preserve the eigenvalues of a density matrix, the global distinguishability is strictly conserved at all times throughout the evolution:
\begin{equation}
\Delta_t^{SE} = \Delta_0^{SE}.
\label{eq:conservation}
\end{equation}
    \item From the subadditivity of the trace distance for initially uncorrelated pair of states \cite{Laine2010}:
    \begin{equation}
        D(\rho_1^S\otimes\rho_1^E,\rho_2^S\otimes\rho_2^E)\leq D(\rho_1^S,\rho_2^S)+D(\rho_1^E,\rho_2^E)
        \label{eq:subadditivity}
    \end{equation}
\end{itemize}

 While the global distinguishability remains conserved, the local distinguishabilities $D_S^t$ and $D_E^t$ fluctuate dynamically as information is exchanged across the partition of the subsystems and hidden within bipartite correlations.
\section{Asymmetric Bounds and the Role of Initial Correlations}\label{sec:initial correlation}

Before establishing our framework, it is useful to examine the asymmetric bounds established in the existing literature, particularly the foundational result of Laine \textit{et al.} \cite{Laine2010}.

If one assumes that the system and environment are initially uncorrelated and that the environment is in a fixed state, $\rho_i^{SE}(0) = \rho_i^S(0) \otimes \rho^E(0)$, the reduced dynamics of $S$ are described by CPTP maps. Under these conditions, the trace distance is monotonically contractive:
\begin{equation}
D_S^t \le D_S^0.
\label{eq:contraction}
\end{equation}
Physically, this implies that the total amount of information flowing back from the environment can never exceed the total amount of information that has flowed out since $t=0$.

However, Laine \textit{et al.} demonstrated that if the initial states contain system-environment correlations, the dynamics can violate Eq.~\eqref{eq:contraction}. By applying the triangle inequality and the data processing inequality to the conserved global trace distance, they derived the bound:
\begin{equation}
D_S^t - D_S^0 \le D(\rho_1^{SE}(0), \rho_2^{SE}(0)) - D_S^0 \equiv I(\rho_1^{SE}, \rho_2^{SE}).
\label{eq:laine_bound}
\end{equation}

The quantity $I(\rho_1^{SE}, \rho_2^{SE})$ represents the distinguishability of the initial global states minus the distinguishability of the initial reduced states. It quantifies the relative information about the initial preparation that lies outside the open system $S$ and is therefore inaccessible to local measurements at $t=0$. Equation \eqref{eq:laine_bound} elegantly proves that the system's trace distance can increase above its initial value, but this net non-Markovian gain is strictly bounded by the information which is initially inaccessible to the open system.

While Eq.~\eqref{eq:laine_bound} represents a major conceptual leap, it only concentrates on what happens to the distinguishability of the system by tracing out the environment, ignoring to keep track on the distinguishability of the environment. Thus, information flow remains a concept which is in one-way track.  

To fully understand the mechanics of memory effects under global unitarity, we should adopt a bipartite approach. We dig deeper into how the shared correlation reservoir distribute non-Markovian memory effects across the entire subsystem partition.

\section{Channel Capacity Limits on State Evolution}\label{sec:prod}

We now transition to our symmetric framework. We begin by establishing the maximum capacity for non-Markovian information backflow in a given subsystem, assuming initially uncorrelated states.

Using eqn.~\eqref{eq:dpi} at any arbitrary time $t$ we get:
\begin{equation}
D_S^t \le \Delta_t^{SE} \quad \text{and} \quad D_E^t \le \Delta_t^{SE}.
\label{eq:data_processing}
\end{equation}

Let us strictly assume that the dynamics begin from uncorrelated initial product states, $\rho_i^{SE}(0) = \rho_i^S(0) \otimes \rho_i^E(0)$. Using eqn.~\eqref{eq:mult} the trace distance of a product state is strictly bounded by the sum of its marginals. Evaluating the initial states yields:
\begin{equation}
\Delta_0^{SE} \le D_S^0 + D_E^0.
\label{eq:initial_bound}
\end{equation}

By utilizing the global conservation law eqn.~\eqref{eq:conservation}, eqn.~\eqref{eq:dpi} and eqn.~\eqref{eq:subadditivity}, we derive:
\begin{align}
0 &= \Delta_0^{SE} - \Delta_t^{SE}\\ \nonumber
&=\tn{\rho_1^{SE}(0)-\rho_2^{SE}(0)}_1-\tn{\rho_1^{SE}(t)-\rho_2^{SE}(t)}_1 \nonumber \\ \nonumber
&=\tn{\rho_1^S(0)\otimes\rho_1^E(0)-\rho_2^S(0)\otimes\rho_2^E(0)}-\tn{\rho_1^{SE}(t)-\rho_2^{SE}(t)}_1 \\
  &\le (D_S^0 + D_E^0) - D_S^t.
\end{align}
Rearranging this inequality to isolate the net change of the system's trace distance yields:
\begin{equation}
D_S^t - D_S^0 \le D_E^0.
\label{eq:sys_bound}
\end{equation}
By symmetry, applying the exact same logic to the environment's data processing bound yields the complementary constraint:
\begin{equation}
D_E^t - D_E^0 \le D_S^0.
\label{eq:env_bound}
\end{equation}

\begin{theorem}
For unitarily evolving initially uncorrelated joint system-environment quantum state,the maximum possible net information gain in one subsystem is rigidly bounded by the initial distinguishability residing in the other subsystem. 
\end{theorem}

Theorem 1 establishes that the local subsystems serve as finite, initial information reservoirs for one another. The system $S$ cannot experience a net increase in trace distance unless the environment $E$ possessed some initial distinguishability ($D_E^0 > 0$) that could be dynamically transferred. 

\section{Dynamics of Bipartite Correlations}\label{sec:mid_corr}

While Theorem 1 effectively bounds the maximum individual gains, it does not fully capture the simultaneous, instantaneous trade-off between the subsystems. To rigorously relate the net changes $\Delta D_S \equiv D_S^t - D_S^0$ and $\Delta D_E \equiv D_E^t - D_E^0$, we must formulate mathematical bounds on their sum.

We add eqn.~\eqref{eq:sys_bound} and \eqref{eq:env_bound} and obtain the absolute upper bound for the joint dynamics:
\begin{equation}
\Delta D_S + \Delta D_E \le (D_S^0 + D_E^0) 
\label{eq:joint_upper bound}
\end{equation}
This upper bound dictates the absolute limit of simultaneous distinguishability gain. 

To establish the corresponding lower bound, we must explicitly account for the dynamically generated bipartite correlations. We decompose the joint state at time $t$ into its marginals and a bipartite correlation operator $\chi_i^{SE}(t)$:
\begin{equation}
\rho_i^{SE}(t) = \rho_i^S(t) \otimes \rho_i^E(t) + \chi_i^{SE}(t).
\end{equation}
By construction, the partial traces of the correlation operators vanish: $\Tr_{S,E}[\chi_i^{SE}(t)] = 0$. We define the correlation difference operator as $\Xi^{SE}(t) \equiv \chi_1^{SE}(t) - \chi_2^{SE}(t)$. 

The exact difference between the two global states at time $t$ is:
\begin{align}
\rho_1^{SE}(t) - \rho_2^{SE}(t) =& \big(\rho_1^S(t) \otimes \rho_1^E(t) - \rho_2^S(t) \otimes \rho_2^E(t)\big) \nonumber \\
&+ \Xi^{SE}(t).
\end{align}
We take the trace norm of both sides and apply the triangle inequality $||A + B||_1 \le ||A||_1 + ||B||_1$. To evaluate the tensor product difference, we add and subtract the cross-term $\rho_2^S(t) \otimes \rho_1^E(t)$, yielding:
\begin{align}
||\rho_1^{SE} - \rho_2^{SE}||_1 \le& ||(\rho_1^S - \rho_2^S) \otimes \rho_1^E||_1 \nonumber \\
&+ ||\rho_2^S \otimes (\rho_1^E - \rho_2^E)||_1 \nonumber \\
&+ ||\Xi^{SE}||_1.
\end{align}
Applying the multiplicativity of the trace norm given as eqn.~\eqref{eq:mult} and dividing by a factor of 2, we arrive at the exact kinematic sum rule:
\begin{equation}
\Delta_0^{SE} \le D_S^t + D_E^t + \frac{1}{2}||\Xi^{SE}(t)||_1.
\label{eq:exact_sum_rule}
\end{equation}

Equation \eqref{eq:exact_sum_rule} is a profound constraint. It demonstrates that the sum of the local distinguishabilities can drop below the global distinguishability only if the difference is absorbed by the correlation norm $\frac{1}{2}||\Xi^{SE}(t)||_1$. 

To isolate the net changes, we subtract $(D_S^0 + D_E^0)$ from both sides of Eq.~\eqref{eq:exact_sum_rule}. We must also account for the geometry of the initial product states. While $\Delta_0^{SE} = D_S^0 + D_E^0$ holds for orthogonal specific cases, the general property is the inequality in Eq.~\eqref{eq:initial_bound}. We define the initial product residual as:
\begin{equation}
\delta_0 \equiv (D_S^0 + D_E^0) - \Delta_0^{SE} \ge 0.
\end{equation}
Substituting this residual into our rearranged sum rule and using eqn. \eqref{eq:joint_upper bound} provides the full bounding picture for subsystem information flow:

\begin{theorem} (Correlation Window). 
For initial product states, the sum of the net changes in subsystem distinguishability is strictly confined to the regime:
\begin{equation}
-\delta_0 - \frac{1}{2}||\Xi^{SE}(t)||_1 \le \Delta D_S + \Delta D_E \le D_S^0 + D_E^0
\label{eq:window}
\end{equation}
\end{theorem}

Theorem 2 completely characterizes the simultaneous trade-off between the subsystems. The right-hand side dictates the maximum collective non-Markovian backflow. The left-hand side dictates that the maximum possible mutual decay (Markovian loss, where both $\Delta D_S$ and $\Delta D_E$ are negative) is governed exactly by the generation of bipartite correlations, alongside any geometric gap present in the initial product states. If both subsystems are losing distinguishability simultaneously, that distinguishability is necessarily being encoded into the correlations represented by $||\Xi^{SE}(t)||_1$.

\section{Impact of Initial Correlations on the Trade-Off Law}\label{sec:initial_corr}

The derivations in Sections IV and V explicitly assumed that the initial states at $t=0$ were uncorrelated product states. We now relax this constraint to generalize our bounds for arbitrary, initially correlated states, synthesizing our symmetric approach with the foundational insights of Laine \textit{et al.} \cite{Laine2010}.

Suppose the initial states themselves contain system-environment correlations:
\begin{equation}
\rho_i^{SE}(0) = \rho_i^S(0) \otimes \rho_i^E(0) + \chi_i^{SE}(0).
\end{equation}
Let $\Xi^{SE}(0) = \chi_1^{SE}(0) - \chi_2^{SE}(0)$ denote the initial correlation difference operator. 

Applying the exact same triangle inequality expansions to the initial state that we applied to derive Eq.~\eqref{eq:exact_sum_rule}, we find that the initial global trace distance is bounded by:
\begin{equation}
\Delta_0^{SE} \le D_S^0 + D_E^0 + \frac{1}{2}||\Xi^{SE}(0)||_1.
\label{eq:initial_corr_bound}
\end{equation}
This replaces the strict product bound of Eq.~\eqref{eq:initial_bound}.

We now substitute this generalized initial bound into the data processing inequality $D_S^t \le \Delta_t^{SE} = \Delta_0^{SE}$. This yields:
\begin{eqnarray}\label{eq:ic_sys}
\nonumber
&&D_S^t \le D_S^0 + D_E^0 + \frac{1}{2}||\Xi^{SE}(0)||_1\\
&\implies&\Delta D_S \le D_E^0 + \frac{1}{2}||\Xi^{SE}(0)||_1
\end{eqnarray}
Similarly, for the environment:
\begin{eqnarray}\label{eq:ic_env}
  \Delta D_E \le D_S^0 + \frac{1}{2}||\Xi^{SE}(0)||_1  
\end{eqnarray}
Adding eqn.~\eqref{eq:ic_sys} and \eqref{eq:ic_env} we arrive at:
\begin{equation}
    \Delta D_S + \Delta D_E \le D_S^0+ D_E^0 +||\Xi^{SE}(0)||_1
\end{equation}
The above equation gives the upper bound of the trade off law for initially correlated states. 

To find a valid lower bound for the sum $\Delta D_S + \Delta D_E$, we first evaluate the kinematics of the initial state at $t=0$. 

The exact difference between the two global states at time $t=0$ is:
\begin{eqnarray}
  \nonumber  
\rho_1^{SE}(0) - \rho_2^{SE}(0) &=& \left( \rho_1^S(0) \otimes \rho_1^E(0) - \rho_2^S(0) \otimes \rho_2^E(0) \right) \\
&&+ \Xi^{SE}(0)
\end{eqnarray}
Taking the trace norm of both sides and applying a variant of the reverse triangle inequality ($||A + B||_1 \ge ||A||_1 - ||B||_1$), we get:
\begin{eqnarray}
\nonumber
||\rho_1^{SE}(0) - \rho_2^{SE}(0)||_1 &\ge& ||\rho_1^S(0) \otimes \rho_1^E(0) - \rho_2^S(0) \otimes \rho_2^E(0)||_1 \\
&&- ||\Xi^{SE}(0)||_1
\end{eqnarray}

Dividing by 2 provides a strict lower bound for the initial global trace distance, $\Delta_0^{SE}$:
\begin{equation}\label{eq:prod}
\Delta_0^{SE} \ge D_{prod}^0 - \frac{1}{2}||\Xi^{SE}(0)||_1
\end{equation}
where $D_{prod}^0$ represents the trace distance of the initial independent product states: $D(\rho_1^S(0) \otimes \rho_1^E(0), \rho_2^S(0) \otimes \rho_2^E(0))$).

Now, we utilize the kinematic sum rule established in eqn.~\eqref{eq:exact_sum_rule} for time $t$:
\begin{equation}
D_S^t + D_E^t \ge \Delta_0^{SE} - \frac{1}{2}||\Xi^{SE}(t)||_1
\end{equation}

To express this inequality in terms of the net change, we subtract the initial local distinguishabilities $(D_S^0 + D_E^0)$ from both sides:
\begin{equation}
\Delta D_S + \Delta D_E \ge \Delta_0^{SE} - (D_S^0 + D_E^0) - \frac{1}{2}||\Xi^{SE}(t)||_1
\end{equation}

Next, we substitute our derived lower bound for $\Delta_0^{SE}$ as in eqn.~\eqref{eq:prod} into this equation:
\begin{eqnarray}
    \nonumber
    \Delta D_S + \Delta D_E \ge \left( D_{prod}^0 
    - \frac{1}{2}||\Xi^{SE}(0)||_1 \right)\\- (D_S^0 + D_E^0) - \frac{1}{2}||\Xi^{SE}(t)||_1
\end{eqnarray}

Rearranging the terms allows us to group the initial state geometry together with the bipartite correlation norms:
\begin{eqnarray}
    \nonumber
    \Delta D_S + \Delta D_E \ge (D_{prod}^0 - D_S^0 - D_E^0)\\ - \frac{1}{2}||\Xi^{SE}(t)||_1 - \frac{1}{2}||\Xi^{SE}(0)||_1
\end{eqnarray}

Based on the subadditivity of the trace distance, we know that the distinguishability of a product state is strictly bounded by the sum of its marginals: $D_{prod}^0 \le D_S^0 + D_E^0$. Therefore, the term $(D_{prod}^0 - D_S^0 - D_E^0)$ is strictly non-positive. We define this initial geometric gap as $-\delta_0$. Thus;
\begin{equation}
\Delta D_S + \Delta D_E \ge -\delta_0 - \frac{1}{2}||\Xi^{SE}(t)||_1 - \frac{1}{2}||\Xi^{SE}(0)||_1
\end{equation}

\begin{theorem} (Generalised Correlation Window). 
    For unitarily evolving joint system-environment quantum states containing initial bipartite correlations, the sum of the net changes in subsystem distinguishability is strictly confined to the generalized regime:
\begin{eqnarray}
\nonumber
    &&-\delta_0-\frac{1}{2}||\Xi^{SE}(t)||_1 - \frac{1}{2}||\Xi^{SE}(0)||_1\\
    &\le& \Delta D_S + \Delta D_E \le D_S^0 + D_E^0 + ||\Xi^{SE}(0)||_1
\end{eqnarray}
\end{theorem}
This theorem unifies the individual subsystem limits of
Theorem~1 and the simultaneous correlation trade-offs of
Theorem~2 into a comprehensive framework for arbitrary
initial states.
The upper bound shows that the maximum collective
non-Markovian backflow is augmented beyond the marginal
reservoir $D_S^0 + D_E^0$ by the initial correlation
difference norm $\|\Xi^{SE}(0)\|_1$, reflecting the
additional capacity supplied by pre-existing bipartite
correlations.
The lower bound is governed by three contributions:
the geometric residual $\delta_0 = D_S^0 + D_E^0 -
D^0_{\mathrm{prod}} \geq 0$ of the initial product of
marginals, the dynamically generated correlation norm
$\frac{1}{2}\|\Xi^{SE}(t)\|_1$, and the initial
correlation norm $\frac{1}{2}\|\Xi^{SE}(0)\|_1$.
In the product-state limit $\|\Xi^{SE}(0)\|_1 = 0$,
where $D^0_{\mathrm{prod}} = \Delta_0^{SE}$, the window
reduces exactly to that of Theorem~2.
For initially correlated states, simultaneous
non-Markovian gain ($\Delta D_S > 0$ and $\Delta D_E >
0$) is physically permissible but is not a free resource:
it requires $\|\Xi^{SE}(0)\|_1 > 0$ and is funded
directly by the net degradation of the initial correlation
reservoir over time.

\section{Numerical Examples}\label{sec:example}
Here, we have used the circuits for creating 4 qubit GHZ state and 4 qubit W state to testify our results. We have taken different pairs of states and computed trace distances. We denote trace distance at the input of the system as $S_{in}$ and of the environment as $E_{in}$. The trace distances at the output is written as $S_{out}$ for the system and $E_{out}$ for the environment. In similar notion, the trace distance of the joint system-environment state at the input of the circuit is denoted as $T_{in}$ and at the output as $T_{out}$.
\begin{enumerate}
    \item Consider a 4-qubit \textbf{GHZ state} 
    $\frac{1}{\sqrt{2}}(\ket{0000} + \ket{1111})$, created from $\ket{0000}$.
    Constructing the circuit :
    \[
    \begin{quantikz}[column sep = 0.45cm, slice all, , slice titles = {$t_{\the\numexpr\col-1\relax}$},slice label style = {inner sep = 0pt}]
    \lstick{$\ket{0}$} & \qw & \gate{H} & \ctrl{1} & \qw      & \qw &\rstick[4]{$\frac{1}{\sqrt{2}}(\ket{0000} + \ket{1111})$}\\
    \lstick{$\ket{0}$} & \qw & \qw      & \targ{}  & \ctrl{1} & \qw &\\
    \lstick{$\ket{0}$} & \qw & \qw      & \qw      & \targ{}  & \ctrl{1} &\\
    \lstick{$\ket{0}$} & \qw & \qw      & \qw      & \qw      & \targ{} &
    \end{quantikz}
    \]

    The evolution of the circuit is decomposed into unitary operations corresponding to time slices $t_i$. All the gates used in the circuit are unitary operators acting on the Hilbert space of the qubits.\\ Consider two orthogonal states $\ket{0010}$ and $\ket{1001}$:
    First, we take the Hadamard gate acting on qubit $0$ between time $t_0$ and $t_1$ i.e $H(0)$ :
    \vspace{-0.3cm}
    \[
    \begin{array}{cc}
    S_{\mathrm{in}} = 1.0, & S_{\mathrm{out}} = 0.999 \\
    E_{\mathrm{in}} = 1.0, & E_{\mathrm{out}} = 0.999 \\
    T_{\mathrm{in}} = 1.0, & T_{\mathrm{out}} = 0.999 \\
    \end{array}
    \]

    Next, we take the controlled-NOT (CNOT) gate acting from qubit $0$ to qubit $1$ between time $t_1$ and $t_2$  i.e CNOT$(0 \rightarrow 1)$ :
    \vspace{-0.3cm}

    \[
    \begin{array}{cc}
    S_{\mathrm{in}} = 0.999, & S_{\mathrm{out}} = 0.999 \\
    E_{\mathrm{in}} = 0.999, & E_{\mathrm{out}} = 0.999 \\
    T_{\mathrm{in}} = 0.999, & T_{\mathrm{out}} = 0.999 \\
    \end{array}
    \]

    Next, we take the controlled-NOT (CNOT) gate acting from qubit $0$ to qubit $2$ between time $t_2$ and $t_3$  i.e CNOT$(0 \rightarrow 2)$ :
    \vspace{-0.1cm}
    \[
    \begin{array}{cc}
    S_{\mathrm{in}} = 0.999, & S_{\mathrm{out}} = 0 \\
    E_{\mathrm{in}} = 0.999, & E_{\mathrm{out}} = 0.999 \\
    T_{\mathrm{in}} = 0.999, & T_{\mathrm{out}} = 0.999 \\
    \end{array}
    \]

    Next, we take the controlled-NOT (CNOT) gate acting from qubit $0$ to qubit $3$ between time $t_3$ and $t_4$  i.e CNOT$(0 \rightarrow 3)$:
     \vspace{-0.1cm}
    \[
    \begin{array}{cc}
    S_{\mathrm{in}} = 0, & S_{\mathrm{out}} = 0 \\
    E_{\mathrm{in}} = 0.999, & E_{\mathrm{out}} = 0 \\
    T_{\mathrm{in}} = 0.999, & T_{\mathrm{out}} = 0.999 \\
    \end{array}
    \]

    Finally, the Total values between time $t_0$ and $t_4$  are given by :
    \[
    \begin{array}{cc}
    S_{\mathrm{in}} = 1.0, & S_{\mathrm{out}} = 0 \\
    E_{\mathrm{in}} = 1.0, & E_{\mathrm{out}} = 0 \\
    T_{\mathrm{in}} = 1.0, & T_{\mathrm{out}} = 1.0\\
    \end{array}
    \]

    Hence, both the system $S$ and the environment $E$ exhibit \textbf{P-divisible} dynamics.

    \item Consider a 4-qubit \textbf{W state} 
    $\frac{1}{2}(\ket{0001} + \ket{0010} + \ket{0100} + \ket{1000})$, created from $\ket{0000}$.
    Constructing the circuit :
    \[
    \begin{quantikz}[column sep = 0.2cm, slice all, , slice titles = {$t_{\the\numexpr\col-1\relax}$},slice label style = {inner sep = 0pt}]
    \lstick{$\ket{0}$} & \qw  & \gate{X} & \gate[wires=2]{PS(\theta_0)}  & \qw      & \qw      & \qw    & \qw &&\\
    \lstick{$\ket{0}$} & \qw   & \qw     &                     & \gate[wires=2]{PS(\theta_1)} & \qw & \gate{Z}  &\qw &&\\
    \lstick{$\ket{0}$} & \qw & \qw   & \qw                &                    & \gate[wires=2]{PS(\theta_2)} & \qw & \qw &&\\
    \lstick{$\ket{0}$} & \qw & \qw      & \qw                & \qw &                    & \qw & \gate{Z} &
    \end{quantikz}
    \]

    where
    $PS_{ij}(\theta)$ denotes the partial swap gate acting on qubits \(i\) and \(j\), defined by
    \[
    PS(\theta)
    =
    \begin{pmatrix}
    1 & 0 & 0 & 0 \\
    0 & \cos\theta & -\sin\theta & 0 \\
    0 & \sin\theta & \cos\theta & 0 \\
    0 & 0 & 0 & 1
    \end{pmatrix},
    \]
    and
    \[
    \theta_0=\cos^{-1}\left(\frac{1}{2}\right),
    \qquad
    \theta_1=\cos^{-1}\left(\frac{1}{\sqrt3}\right),
    \]
    \[
    \theta_2=\cos^{-1}\left(\frac{1}{\sqrt2}\right)
    \]

    The evolution of the circuit is decomposed into unitary operations corresponding to time slices $t_i$. All the gates used in the circuit are unitary operators acting on the Hilbert space of the qubits.\\
    \renewcommand{\labelenumi}{(\roman{enumi})}
    \setcounter{enumi}{0}
    \item Consider two states $\ket{0011}$ and $1/\sqrt{2}\ket{0001} + 1/\sqrt{2}\ket{0110}$:
    First, we take the Pauli-$X$ gate acting on qubit $0$ between time $t_0$ and $t_1$ i.e $X(0)$ :
    \vspace{-0.3cm}
    \[
    \begin{array}{cc}
    S_{\mathrm{in}} = 0.5, & S_{\mathrm{out}} = 0.5 \\
    E_{\mathrm{in}} = 1.0, & E_{\mathrm{out}} = 1.0 \\
    T_{\mathrm{in}} = 1.0, & T_{\mathrm{out}} = 1.0 \\
    \end{array}
    \]

    Next, we take the Partial Swap ($\theta$) gate acting from qubit $0$ to qubit $1$ between time $t_1$ and $t_2$ i.e PS$(0 \rightarrow 1)$ :
    \vspace{-0.1cm}

    \[
    \begin{array}{cc}
    S_{\mathrm{in}} = 0.5, & S_{\mathrm{out}} = 0.5 \\
    E_{\mathrm{in}} = 1.0, & E_{\mathrm{out}} = 1.0 \\
    T_{\mathrm{in}} = 1.0, & T_{\mathrm{out}} = 1.0 \\
    \end{array}
    \]

    Next, we take the Partial Swap ($\theta$) gate acting from qubit $0$ to qubit $2$ between time $t_2$ and $t_3$  i.e PS$(1 \rightarrow 2)$ :
    \vspace{-0.1cm}
    \[
    \begin{array}{cc}
    S_{\mathrm{in}} = 0.5, & S_{\mathrm{out}} = 0.6476 \\
    E_{\mathrm{in}} = 1.0, & E_{\mathrm{out}} = 0.5833 \\
    T_{\mathrm{in}} = 1.0, & T_{\mathrm{out}} = 1.0 \\
    \end{array}
    \]

    Next, we take the Partial Swap ($\theta$) gate acting from qubit $0$ to qubit $3$ between time $t_3$ and $t_4$  i.e PS$(2 \rightarrow 3)$:
     \vspace{-0.1cm}
    \[
    \begin{array}{cc}
    S_{\mathrm{in}} = 0.6476, & S_{\mathrm{out}} = 0.6476 \\
    E_{\mathrm{in}} = 0.5833, & E_{\mathrm{out}} = 0.5833 \\
    T_{\mathrm{in}} = 1.0, & T_{\mathrm{out}} = 1.0 \\
    \end{array}
    \]

    Next, we take the Pauli-$Z$ gate acting 0n qubit $1$ between time $t_4$ and $t_5$ i.e Z$(1)$:
     \vspace{-0.1cm}
    \[
    \begin{array}{cc}
    S_{\mathrm{in}} = 0.6476, & S_{\mathrm{out}} = 0.6476 \\
    E_{\mathrm{in}} = 0.5833, & E_{\mathrm{out}} = 0.5833 \\
    T_{\mathrm{in}} = 1.0, & T_{\mathrm{out}} = 1.0 \\
    \end{array}
    \]

    Next, we take the Pauli-$Z$ gate acting 0n qubit $3$ between time $t_5$ and $t_6$ i.e Z$(3)$:
     \vspace{-0.1cm}
    \[
    \begin{array}{cc}
    S_{\mathrm{in}} = 0.6476, & S_{\mathrm{out}} = 0.6476 \\
    E_{\mathrm{in}} = 0.5833, & E_{\mathrm{out}} = 0.5833 \\
    T_{\mathrm{in}} = 1.0, & T_{\mathrm{out}} = 1.0 \\
    \end{array}
    \]

    Finally, the Total values between time $t_0$ and $t_6$ are given by :
    \[
    \begin{array}{cc}
    S_{\mathrm{in}} = 0.5, & S_{\mathrm{out}} = 0.6476\\
    E_{\mathrm{in}} = 1.0, & E_{\mathrm{out}} = 0.5833 \\
    T_{\mathrm{in}} = 1.0, & T_{\mathrm{out}} = 1.0\\
    \end{array}
    \]

    Hence, the system $S$ is \textbf{P-indivisible} and the environment $E$ exhibit \textbf{P-divisible} dynamics.

    \setcounter{enumi}{1}
    \item Consider two states $\ket{1100}$ and $1/\sqrt{2}\ket{0100} + 1/\sqrt{2}\ket{1001}$:

    First, we take the Pauli-$X$ gate acting on qubit $0$ between time $t_0$ and $t_1$ i.e $X(0)$ :
    \vspace{-0.3cm}
    \[
    \begin{array}{cc}
    S_{\mathrm{in}} = 1.0, & S_{\mathrm{out}} = 1.0 \\
    E_{\mathrm{in}} = 0.5, & E_{\mathrm{out}} = 0.5 \\
    T_{\mathrm{in}} = 1.0, & T_{\mathrm{out}} = 1.0 \\
    \end{array}
    \]

    Next, we take the Partial Swap ($\theta$) gate acting from qubit $0$ to qubit $1$ between time $t_1$ and $t_2$ i.e PS$(0 \rightarrow 1)$ :
    \vspace{-0.3cm}

    \[
    \begin{array}{cc}
    S_{\mathrm{in}} = 1.0, & S_{\mathrm{out}} = 1.0 \\
    E_{\mathrm{in}} = 0.5, & E_{\mathrm{out}} = 0.5 \\
    T_{\mathrm{in}} = 1.0, & T_{\mathrm{out}} = 1.0 \\
    \end{array}
    \]

    Next, we take the Partial Swap ($\theta$) gate acting from qubit $0$ to qubit $2$ between time $t_2$ and $t_3$ i.e PS$(1 \rightarrow 2)$ :
    \vspace{-0.1cm}
    \[
    \begin{array}{cc}
    S_{\mathrm{in}} = 1.0, & S_{\mathrm{out}} = 0.55046 \\
    E_{\mathrm{in}} = 0.5, & E_{\mathrm{out}} = 0.66667 \\
    T_{\mathrm{in}} = 1.0, & T_{\mathrm{out}} = 1.0 \\
    \end{array}
    \]

    Next, we take the Partial Swap ($\theta$) gate acting from qubit $0$ to qubit $3$ between time $t_3$ and $t_4$ i.e PS$(2 \rightarrow 3)$:
     \vspace{-0.1cm}
    \[
    \begin{array}{cc}
    S_{\mathrm{in}} = 0.55046, & S_{\mathrm{out}} = 0.55046 \\
    E_{\mathrm{in}} = 0.66667, & E_{\mathrm{out}} = 0.66667 \\
    T_{\mathrm{in}} = 1.0, & T_{\mathrm{out}} = 1.0 \\
    \end{array}
    \]

    Next, we take the Pauli-$Z$ gate acting 0n qubit $1$ between time $t_4$ and $t_5$ i.e Z$(1)$:
     \vspace{-0.1cm}
    \[
    \begin{array}{cc}
    S_{\mathrm{in}} = 0.55046, & S_{\mathrm{out}} = 0.55046 \\
    E_{\mathrm{in}} = 0.66667, & E_{\mathrm{out}} = 0.66667 \\
    T_{\mathrm{in}} = 1.0, & T_{\mathrm{out}} = 1.0 \\
    \end{array}
    \]

    Next, we take the Pauli-$Z$ gate acting 0n qubit $3$ between time $t_5$ and $t_6$ i.e Z$(3)$:
     \vspace{-0.1cm}
    \[
    \begin{array}{cc}
    S_{\mathrm{in}} = 0.55046, & S_{\mathrm{out}} = 0.55046 \\
    E_{\mathrm{in}} = 0.66667, & E_{\mathrm{out}} = 0.66667 \\
    T_{\mathrm{in}} = 1.0, & T_{\mathrm{out}} = 1.0 \\
    \end{array}
    \]

    Finally, the Total values between time $t_0$ and $t_6$ are given by :
    \[
    \begin{array}{cc}
    S_{\mathrm{in}} = 1.0, & S_{\mathrm{out}} = 0.55046\\
    E_{\mathrm{in}} = 0.5, & E_{\mathrm{out}} = 0.66667 \\
    T_{\mathrm{in}} = 1.0, & T_{\mathrm{out}} = 1.0\\
    \end{array}
    \]

    Hence, the system $S$ is \textbf{P-divisible} and the environment $E$ exhibit \textbf{P-indivisible} dynamics.

    \setcounter{enumi}{2}
    \item Consider two states $1/2\ket{0000} + 1/2\ket{1111} + 1/2\ket{1000} + 1/2\ket{0001} $ and $1/\sqrt{6}\ket{0000} + 1/\sqrt{6}\ket{1111} + 1/\sqrt{6}\ket{0100} + 1/\sqrt{6}\ket{0010} + 1/\sqrt{6}\ket{1011} + 1/\sqrt{6}\ket{1101}$:

    First, we take the Pauli-$X$ gate acting on qubit $0$ between time $t_0$ and $t_1$ i.e $X(0)$ :
    \vspace{-0.3cm}
    \[
    \begin{array}{cc}
    S_{\mathrm{in}} = 0.47513, & S_{\mathrm{out}} = 0.47513 \\
    E_{\mathrm{in}} = 0.47513, & E_{\mathrm{out}} = 0.47513 \\
    T_{\mathrm{in}} = 0.91287, & T_{\mathrm{out}} = 0.91287 \\
    \end{array}
    \]

    Next, we take the Partial Swap ($\theta$) gate acting from qubit $0$ to qubit $1$ between time $t_1$ and $t_2$ i.e PS$(0 \rightarrow 1)$ :
    \vspace{-0.3cm}

    \[
    \begin{array}{cc}
    S_{\mathrm{in}} = 0.47513, & S_{\mathrm{out}} = 0.47513 \\
    E_{\mathrm{in}} = 0.47513, & E_{\mathrm{out}} = 0.47513 \\
    T_{\mathrm{in}} = 0.91287, & T_{\mathrm{out}} = 0.91287 \\
    \end{array}
    \]

    Next, we take the Partial Swap ($\theta$) gate acting from qubit $0$ to qubit $2$ between time $t_2$ and $t_3$ i.e PS$(1 \rightarrow 2)$ :
    \vspace{-0.1cm}
    \[
    \begin{array}{cc}
    S_{\mathrm{in}} = 0.47513, & S_{\mathrm{out}} = 0.62773 \\
    E_{\mathrm{in}} = 0.47513, & E_{\mathrm{out}} = 0.55131 \\
    T_{\mathrm{in}} = 0.91287, & T_{\mathrm{out}} = 0.91287 \\
    \end{array}
    \]

    Next, we take the Partial Swap ($\theta$) gate acting from qubit $0$ to qubit $3$ between time $t_3$ and $t_4$ i.e PS$(2 \rightarrow 3)$:
     \vspace{-0.1cm}
    \[
    \begin{array}{cc}
    S_{\mathrm{in}} = 0.62773, & S_{\mathrm{out}} = 0.62773 \\
    E_{\mathrm{in}} = 0.55131, & E_{\mathrm{out}} = 0.55131 \\
    T_{\mathrm{in}} = 0.91287, & T_{\mathrm{out}} = 0.91287 \\
    \end{array}
    \]

    Next, we take the Pauli-$Z$ gate acting 0n qubit $1$ between time $t_4$ and $t_5$ i.e Z$(1)$:
     \vspace{-0.1cm}
    \[
    \begin{array}{cc}
    S_{\mathrm{in}} = 0.62773, & S_{\mathrm{out}} = 0.62773 \\
    E_{\mathrm{in}} = 0.55131, & E_{\mathrm{out}} = 0.55131 \\
    T_{\mathrm{in}} = 0.91287, & T_{\mathrm{out}} = 0.91287 \\
    \end{array}
    \]

    Next, we take the Pauli-$Z$ gate acting 0n qubit $3$ between time $t_5$ and $t_6$ i.e Z$(3)$:
     \vspace{-0.1cm}
    \[
    \begin{array}{cc}
    S_{\mathrm{in}} = 0.62773, & S_{\mathrm{out}} = 0.62773 \\
    E_{\mathrm{in}} = 0.55131, & E_{\mathrm{out}} = 0.55131 \\
    T_{\mathrm{in}} = 0.91287, & T_{\mathrm{out}} = 0.91287 \\
    \end{array}
    \]

    Finally, the Total values between time $t_0$ and $t_6$ are given by :
    \[
    \begin{array}{cc}
    S_{\mathrm{in}} = 0.47513, & S_{\mathrm{out}} = 0.62773 \\
    E_{\mathrm{in}} = 0.47513, & E_{\mathrm{out}} = 0.55131 \\
    T_{\mathrm{in}} = 0.91287, & T_{\mathrm{out}} = 0.91287 \\
    \end{array}
    \]

    Hence, both the system $S$ and the environment $E$ exhibit \textbf{P-indivisible} dynamics.

\end{enumerate}

\section{ Discussion}\label{sec:discussion}

We now verify the bounds of Theorems~1--3 against the
numerical data.

\noindent\textbf{a. GHZ circuit.}
Both subsystems exhibit P-divisible dynamics, with
$\Delta D_S = \Delta D_E = -1.0$.
Theorem~1 is satisfied with maximal slack:
the net change in each subsystem is negative and well
within the initial reservoir of its counterpart.
The Correlation Window of Theorem~2 requires
$-\delta_0 - \frac{1}{2}\|\Xi^{SE}(t)\|_1 \leq -2.0$,
which, with $\delta_0 = 1.0$, forces $\|\Xi^{SE}(t)\|_1 \geq 2.0$.
This is the phenomenon of \emph{subsystem locking}: the
global distinguishability is perfectly conserved
($T_{\mathrm{in}} = T_{\mathrm{out}} = 1.0$) while both
local distinguishabilities vanish completely, proving that
the entire combined loss of $2.0$ units is irreversibly
absorbed into the dynamically generated bipartite
correlation reservoir.
Since $\|\Xi^{SE}(0)\|_1 = 0$ for these product input
states, Theorem~3 reduces exactly to Theorem~2.

\noindent\textbf{b. W circuit --- system backflow [Example 2(i)].} The system gains distinguishability ($\Delta D_S = +0.1476$, P-indivisible) while the environment loses it ($\Delta D_E = -0.4167$, P-divisible). Because the second input state, $1/\sqrt{2}(|0001\rangle + |0110\rangle)$, contains bipartite correlations, the initial correlation difference norm is non-zero ($||\Xi^{SE}(0)||_1 > 0$). Therefore, the dynamics are governed by the generalized bounds of Theorem 3. The system's non-Markovian gain is safely accommodated within the modified individual bound of eqn.~\eqref{eq:ic_sys}: $\Delta D_S = 0.1476 \le D_E^0 + \frac{1}{2}||\Xi^{SE}(0)||_1$, funded by the environment's initial distinguishability reservoir and the initial correlations. The collective sum of the net changes, $\Delta D_S + \Delta D_E = -0.2691$, easily satisfies the augmented upper bound of Theorem 3, $-0.2691 \le 1.5 + ||\Xi^{SE}(0)||_1$. The net change of $-0.2691$ is well above the negative floor set by $-\delta_0=-0.5$ and the negative of the correlation norms, confirming that the mutual trade-off adheres strictly to the capacity limits of the initial correlated states.

\noindent\textbf{c. W circuit --- environment backflow [Example 2(ii)].} This example is the symmetric counterpart of Example 2(i) under the interchange of the S and E labels. The environment gains distinguishability ($\Delta D_E = +0.16667$, P-indivisible) and the system loses it ($\Delta D_S = -0.44954$, P-divisible). Similarly, the second input state $1/\sqrt{2}(|0100\rangle + |1001\rangle)$ introduces initial bipartite correlations ($||\Xi^{SE}(0)||_1 > 0$), which requires the use of Theorem 3. The gain of the environment successfully satisfies the individual bound of eqn.~\eqref{eq:ic_env}: $\Delta D_E = 0.16667 \le D_S^0 + \frac{1}{2}||\Xi^{SE}(0)||_1$. The collective sum $\Delta D_S + \Delta D_E = -0.28287$ is structurally comparable to $-0.2691$ of Example 2(i) and comfortably satisfies the upper bound of Theorem 3 ($-0.28287 \le 1.5 + ||\Xi^{SE}(0)||_1$), as well as the lower bound of $-\delta_0-\frac{1}{2}||\Xi^{SE}(t)||_1 - \frac{1}{2}||\Xi^{SE}(0)||_1$, where $-\delta_0=-0.5$. 


\noindent\textbf{d. W circuit — initially correlated states [Example~2(iii)].}
Both subsystems exhibit simultaneous P-indivisible
dynamics: $\Delta D_S = +0.15260$ and $\Delta D_E =
+0.07618$, giving a positive collective sum
$\Delta D_S + \Delta D_E = +0.22878$.
Both input states are entangled joint states, with
$\Delta_0^{SE} = 0.91287 < D_S^0 + D_E^0 = 0.95026$
confirming $\|\Xi^{SE}(0)\|_1 > 0$.
While the individual Theorem~1 bounds are numerically
satisfied, Theorem~1 alone cannot account for the
simultaneous positivity of both $\Delta D_S$ and
$\Delta D_E$: for product initial states, each
subsystem's gain would require the other to act as a
simultaneous donor, which is kinematically inconsistent.
Theorem~3 resolves this: with $\delta_0 = D_S^0 + D_E^0
- D^0_{\rm prod} \geq 0$ and $\|\Xi^{SE}(0)\|_1 > 0$,
the generalized upper bound
$0.22878 \leq 0.95026 + \|\Xi^{SE}(0)\|_1$ is trivially
satisfied, and the lower bound
$-\delta_0 - \frac{1}{2}\|\Xi^{SE}(t)\|_1 -
\frac{1}{2}\|\Xi^{SE}(0)\|_1 \leq 0 < +0.22878$
is also satisfied.
Most importantly, the positive collective gain directly
implies $\|\Xi^{SE}(0)\|_1 > \|\Xi^{SE}(t)\|_1$:
the initial bipartite correlations are consumed as a
resource to fund simultaneous P-indivisibility in both
subsystems, confirming the central physical claim of
Theorem~3.

\section{Conclusion}\label{sec:conclusion}

We have developed a symmetric kinematic framework for
P-divisibility in bipartite open quantum systems.
The central anchor is the conservation of the global trace
distance under unitary evolution, which we use to derive
tight bounds on how distinguishability is redistributed
between the system, the environment, and the dynamically
generated correlation reservoir.
 
Theorem~1 establishes that for initially uncorrelated
states the net non-Markovian gain in one subsystem cannot
exceed the initial distinguishability of the other.
The two subsystems thus act as finite, mutually exclusive
information reservoirs, and P-indivisibility in one
requires a non-trivial initial endowment in the other.
This result symmetrizes and extends the asymmetric bound
of Laine et al.~\cite{Laine2010}.
 
Theorem~2 extends the picture to the simultaneous dynamics
of both subsystems, confining $\Delta D_S + \Delta D_E$
within a two-sided Correlation Window.
The upper bound caps the maximum collective non-Markovian
backflow at $D_S^0 + D_E^0$.
The lower bound is the physically richer result: any
simultaneous loss of distinguishability across the
partition must be exactly absorbed by the growth of the
bipartite correlation norm $\|\Xi^{SE}(t)\|_1$. 
 
Theorem~3 generalizes the framework to initially correlated
states.
The initial correlation difference norm $\|\Xi^{SE}(0)\|_1$
augments the upper capacity bound and expands the lower
bound of the window, making simultaneous P-indivisibility
of both $S$ and $E$ physically permissible.
Such simultaneous backflow is not a free resource: it is
funded by the degradation of $\|\Xi^{SE}(0)\|_1$ over
time, establishing initial correlations as a consumable
dynamical resource for non-Markovian gain.
Theorem~3 recovers Theorem~2 exactly in the product-state
limit.
 
The GHZ circuit example demonstrates subsystem locking,
where the lower bound of Theorem~2 forces the inference
$\|\Xi^{SE}(t)\|_1 \geq 2.0$ from local data alone.
The W circuit examples with one product state and another correlated state satisfies Theorem~3 in asymmetric backflow directions.
The W circuit example with initially correlated states
demonstrates simultaneous P-indivisibility, explicable through Theorem~3.
 
These symmetric bounds provide a complete kinematic
accounting of non-Markovian information flow under global
unitary dynamics, and may find application in the design
of memory-assisted quantum information protocols where
engineered initial correlations serve as a resource for
enhancing local non-Markovian capacity.

\bibliographystyle{unsrt}
\bibliography{positive}
\end{document}